\documentclass[5p,times]{elsarticle}

\usepackage{graphicx}
\usepackage{amsmath}
\usepackage{siunitx}

\journal{Nuclear Instruments and Methods in Physics Research A}

\title{Recent advances and trends in pattern recognition and data analysis for RICH detectors}
\author[ULJ,IJS]{L.~\v{S}antelj}  
\ead{luka.santelj@ijs.si}

\address[ULJ]{University of Ljubljana, Slovenia}
\address[IJS]{Jo\v{z}ef Stefan Institute, Ljubljana, Slovenia}

\begin{document}
\sloppy

\begin{abstract}
Ring Imaging Cherenkov (RICH) detectors are a key component of particle identification systems in many particle, nuclear and astroparticle physics experiments. Their ultimate performance depends not only on detector design and hardware implementation, but also crucially on the quality of pattern recognition and data analysis algorithms used to reconstruct Cherenkov ring images and to perform particle identification. In recent years, significant advances have been made both in traditional reconstruction approaches, such as likelihood-based methods and Hough-transform techniques, and in the application of modern machine learning tools. This contribution reviews the current state of RICH reconstruction algorithms, highlights representative use cases from operating experiments, and discusses emerging trends including global particle identification strategies and generative machine learning approaches for fast simulation and reconstruction.
\end{abstract}


\maketitle

\section{Introduction}
Reliable particle identification (PID) is an essential ingredient in a wide range of particle, nuclear and astroparticle physics experiments. The ability to separate particle species over a broad momentum range enables precise measurements, background suppression, and efficient use of collected data. Ring Imaging Cherenkov detectors are among the most powerful PID devices, providing excellent separation power through reconstruction of Cherenkov photon emission patterns.

Achieving optimal PID performance requires both high-quality detector hardware and sophisticated data reconstruction algorithms. While detector upgrades are often costly and limited in frequency, reconstruction algorithms can be continuously improved throughout the lifetime of an experiment, and even retrospectively applied to already recorded data. The development of efficient and robust reconstruction methods is therefore a critical and cost-effective way to fully exploit the physics potential of existing experimental apparatuses.

In this contribution, we provide an overview of pattern recognition and data analysis techniques used in RICH detectors, with an emphasis on recent developments and emerging trends. We first review established reconstruction approaches, including likelihood-based methods, global event-level strategies, and alternative pattern-recognition techniques such as Hough transforms, which continue to form the backbone of RICH data analysis (Section~2). We then discuss the increasing role of machine learning in particle identification and ring reconstruction, highlighting both performance gains and practical challenges related to robustness and systematic uncertainties (Section~3). Finally, we summarize recent progress in generative machine learning models for fast simulation and reconstruction of Cherenkov detectors, and outline their potential impact on future experiments and large-scale data analysis workflows (Section~4).

\section{Traditional approaches to RICH reconstruction}

\begin{figure}[!ht]
    \centering
    \includegraphics[width=0.7\linewidth]{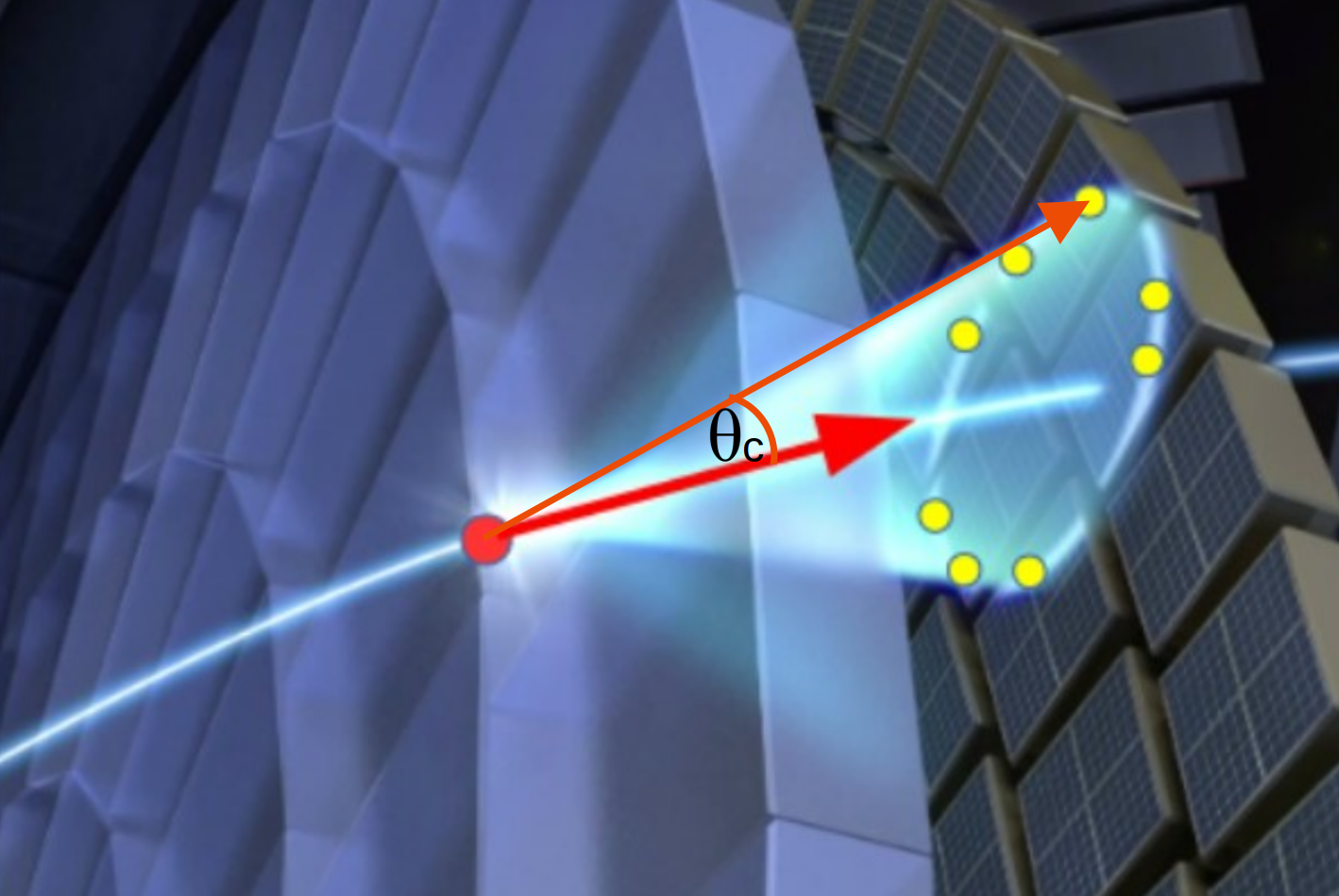}
    \caption{Illustration of the Cherenkov ring emission and Cherenkov angle definition in the typical geometrical configuration of proximity focusing RICH detector. The yellow dots illustrate detected Cherenkov photons.}
    \label{fig:sketch}
\end{figure}

The most elementary approach to Cherenkov-based particle identification, in experiments where the track parameters are provided by other detector subsystems, consists of reconstructing, for each detected photon, the Cherenkov emission angle based on the known detector geometry and the reconstructed track trajectory. Using the assumed photon emission point along the track and the measured hit position on the photon detector plane, the Cherenkov angle $\theta_c$ can be inferred through geometrical relations or ray-tracing procedures. For a proximity focusing RICH with thin radiator layer the configuration is illustrated on Figure \ref{fig:sketch}. The reconstructed single-photon angles are then compared to the expected Cherenkov angle $\theta_c^{h}$ for a given particle hypothesis $h$, determined from the particle momentum and the refractive index of the radiator. In its simplest form, particle identification can be performed by evaluating the compatibility of the reconstructed angle distribution with the expected value for each hypothesis, for example by constructing a $\chi^2$-like estimator based on the deviations $\theta_c - \theta_c^{h}$.

\subsection{Likelihood-based methods}
Likelihood-based reconstruction is the most widely used approach in RICH detectors when external track information is available. Track parameters, such as position, direction, and momentum in the radiator volume, are typically provided by other detector subsystems and serve as input to the RICH reconstruction.

For a given particle hypothesis $h$ (e.g. $e$, $\mu$, $\pi$, $K$, $p$), the expected pattern of detected Cherenkov photons is compared to the observed one \cite{likeli}. The likelihood is constructed as a product over detector pixels,
\begin{equation}
\mathcal{L}^h = \prod_i p_i^h,
\label{eq:like}
\end{equation}
where $p_i^h$ denotes the probability of observing the measured number of photons $m_i$ on pixel $i$, given the expected number $n_i^h$ for hypothesis $h$. For Poisson-distributed photon counts, this probability can be written as
\begin{equation}
p_i^h = \frac{(n_i^h)^{m_i}}{m_i!} e^{-n_i^h}.
\end{equation}
The central challenge of the likelihood method is the accurate and efficient evaluation of the expected photon yields $n_i^h$.

In detectors with binary photon detection, where the readout records only whether a pixel is fired or not, the likelihood can be expressed in a compact logarithmic form. 
Assuming Poisson statistics for the photon multiplicity, the probability for a pixel not to fire is $e^{-n_i^{h}}$, while the probability to fire is $1-e^{-n_i^{h}}$. If $\mathcal{F}$ denotes the set of all fired pixels, the logarithm of the likelihood (\ref{eq:like}) can be re-written as
\begin{equation}
\ln \mathcal{L}_{\mathrm{bin}}^{h}
= \sum_{i\in\mathcal{F}} \ln\!\left(1-e^{-n_i^{h}}\right) - N^{h},
\label{eq:logL_binary}
\end{equation}
where $N^{h}=\sum_i n_i^{h}$ is the total expected number of detected photons for hypothesis $h$. In low-occupancy conditions, this expression is particularly efficient to evaluate, as the first term requires summation over fired pixels only, while the contribution of non-fired pixels is fully contained in the global expectation value $N^{h}$.

\subsubsection{Evaluation of expected photon yields}
The expected photon yield on a given pixel depends on detector geometry, optical properties, track parameters, and the assumed particle hypothesis. For each detected hit, the Cherenkov angles $(\theta_c, \phi_c)$ can be reconstructed under the assumption of a photon emission point along the particle trajectory in the radiator. This corresponds to a reverse ray-tracing problem, which can be solved analytically, semi-analytically, or numerically depending on the detector complexity and required precision.

Signal contributions from unscattered photons are typically described by a probability density function (PDF) in Cherenkov angle space, often approximated by a Gaussian distribution around the expected Cherenkov angle for hypothesis $h$. The width of the distribution represents the Cherenkov angle resolution which typically receives contributions from the track parameter uncertainties, assumed photon emission position in the radiator (due to its finite thickness) and pixelization of the photon-detector plane. Background contributions, originating from scattered photons, secondary particles, or detector noise, are incorporated through additional components of the PDF.

\begin{figure}[!ht]
    \centering
    \includegraphics[width=1\linewidth]{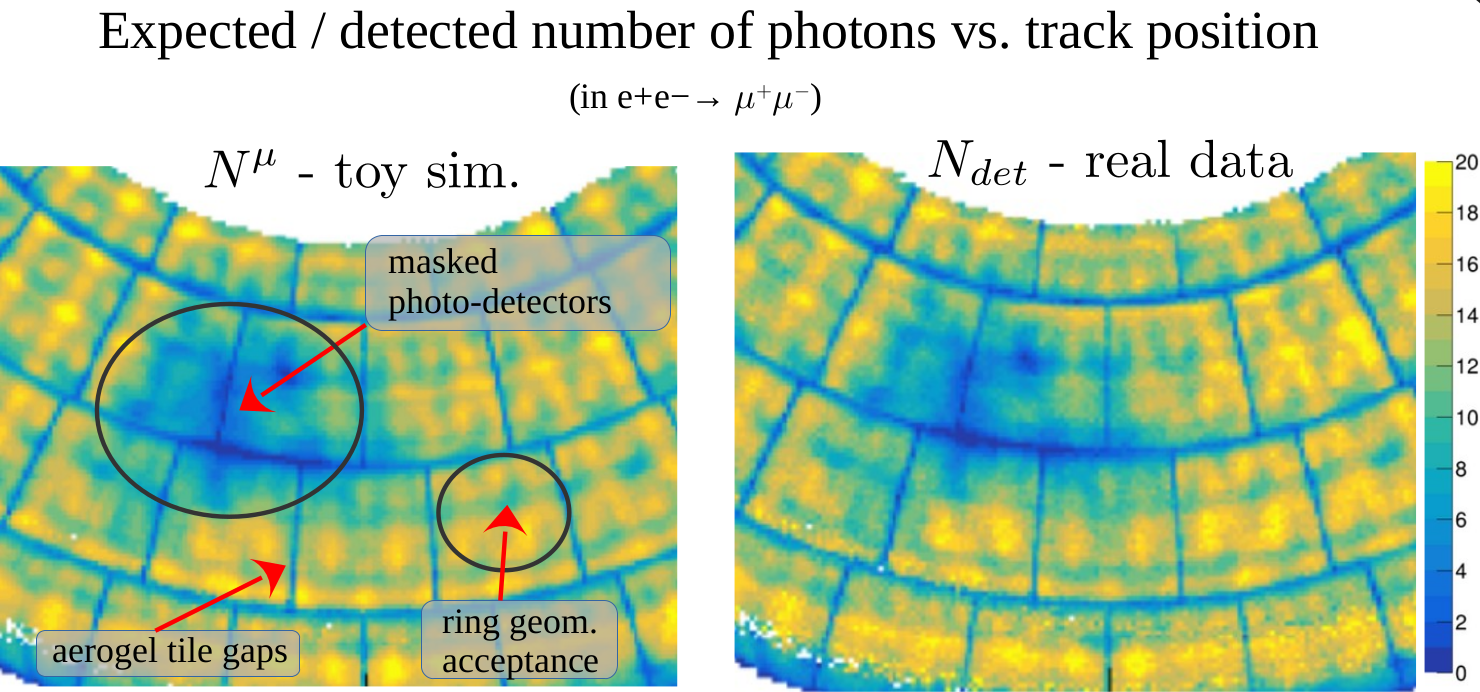}
    \caption{Comparison of the expected number of detected photons for high-energy muons (left) with the corresponding measured number in data (right) in Belle II aerogel RICH, depending on the particle impact position on the aerogel radiator plane. $N^\mu$ is evaluated using toy simulation on track-by-track basis within the likelihood evaluation \cite{belle2_arich1}.}
    \label{fig:arich_N}
\end{figure}

The total expected number of photons $N^h$ can be estimated from the theoretical photon yield corrected for detector efficiencies and average geometrical acceptance. In practice, many experiments rely on fast ray-tracing ``toy simulations'' performed on a track-by-track basis to obtain a more accurate estimate of $N^h$. 
Figure \ref{fig:arich_N} compares the expected number of detected photons for high-energy muons, $N^\mu$, with the corresponding measured number $N_{\mathrm{det}}$ in data for the aerogel RICH detector of Belle~II \cite{belle2_arich1}, shown as a function of the particle incidence position on the aerogel radiator plane. A very good agreement between expectation and measurement is observed across the detector acceptance, while several detector-related effects influencing $N_\mu$ are clearly visible.

\subsubsection{PDF projection to the detector plane}

In geometry configurations where the photon-detector pixel size in Cherenkov space is not small compared to the intrinsic scale of the PDF features, such as the single–photon Cherenkov angle resolution, an accurate evaluation of $n^h_i$ requires integrating the PDF over the pixel surface. Performing this integration directly in Cherenkov angle space is generally non-trivial, as the mapping of a detector pixel to Cherenkov coordinates depends on the optical geometry, photon emission point, and track parameters. In such cases projecting the PDF from the Cherenkov space onto the photon detector plane can be beneficial.  Working in the physical detector plane allows the PDF to be integrated naturally over the pixel area, avoiding complicated coordinate transformations. In addition, several contributions to the PDF, such as optical distortions, detector acceptance effects, and background components, are more conveniently and realistically described in the detector plane than in Cherenkov space. This of course comes at a cost of carrying out the PDF projection. 

\begin{figure}[!ht]
    \centering
    \includegraphics[width=0.9\linewidth]{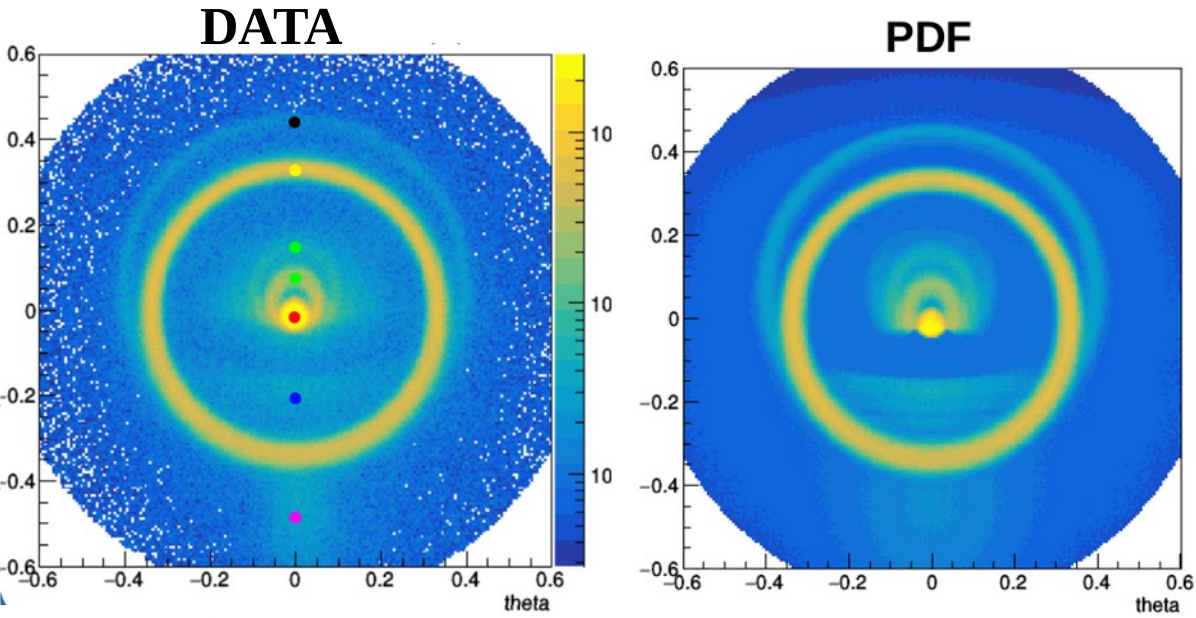}
    \caption{Comparison of Cherenkov ring image for high-energy muons in measured data (accumulated over $\sim 10$k tracks) and the PDF used for muon hypothesis likelihood evaluation. Both pictures are shown in the Cherenkov space. Several detailed effects of ring image mainly originating from photon reflections within the photon detectors can be seen on both pictures \cite{belle2_arich2}.}
    \label{fig:arich_pdf}
\end{figure}

This approach is adopted e.g. in Belle II aerogel RICH reconstruction \cite{belle2_arich2}, where at the first step the Cherenkov angle $(\theta_c, \phi_c)$ is reconstructed assuming that the photon originates from the mean emission point in the aerogel. This is then followed by propagation of a ``toy'' photon from the same point in the aerogel in the direction expected for hypothesis $h$ $(\theta^h_c, \phi_c)$. Around the resulting impact point on the detector plane, the expected photon distribution is modeled by a Gaussian profile in the radial direction of the ring and an a flat distribution in the azimuthal direction (as the pixel size is small compared to the typical Cherenkov ring radius this is good approximation). The finite pixel size is taken into account by integrating the PDF over the pixel surface. The width of the Gaussian is evaluated on a track-by-track basis and incorporates effects the photon emission point uncertainty and the resolution of position and direction of particle on the radiator plane (given by the Belle II tracking system). Several background components are also included in the PDF, such as e.g. Cherenkov photons originating from the particle impacting the quartz window of photon detectors. Figure \ref{fig:arich_pdf} shows comparison of Cherenkov ring for high-energy muons in data (accumulated over $\sim 10000$ tracks) with the PDF used to evaluate muon hypothesis likelihood. Note that the PDF depends on track parameters and hypothesis and is constructed on track-by-track basis. 

\subsection{Global likelihood methods}

Global likelihood methods are particularly important in environments with high track multiplicities, where Cherenkov rings from different particles may overlap significantly on the photon detector plane. In such conditions, assigning photon hits to individual tracks in a purely local, track-by-track manner can lead to sub-optimal particle identification performance. Instead, the event likelihood is constructed from the combined contributions of all reconstructed tracks under a given set of particle hypotheses, allowing correlations between overlapping ring patterns to be taken into account.

\begin{figure}[!ht]
    \centering
    \includegraphics[width=0.45\linewidth]{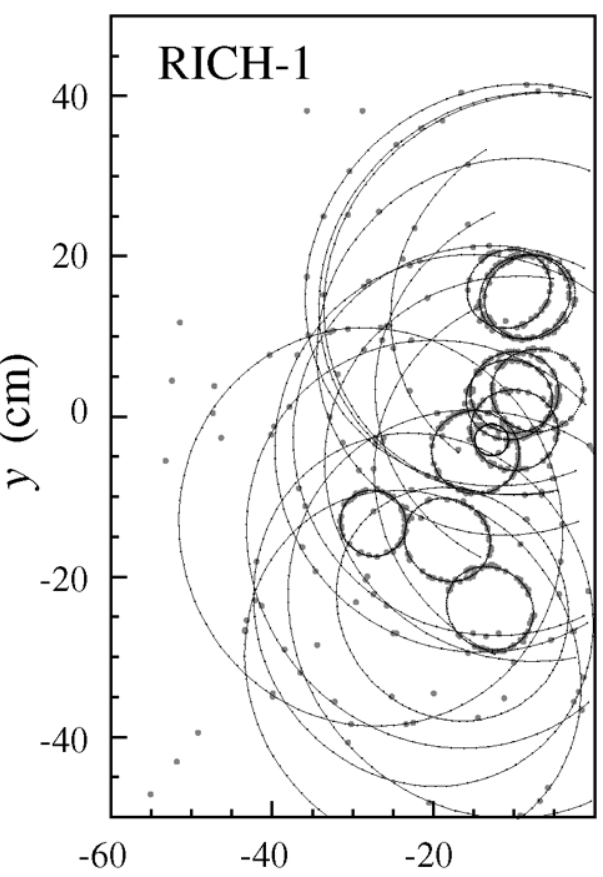}
    \includegraphics[width=0.54\linewidth]{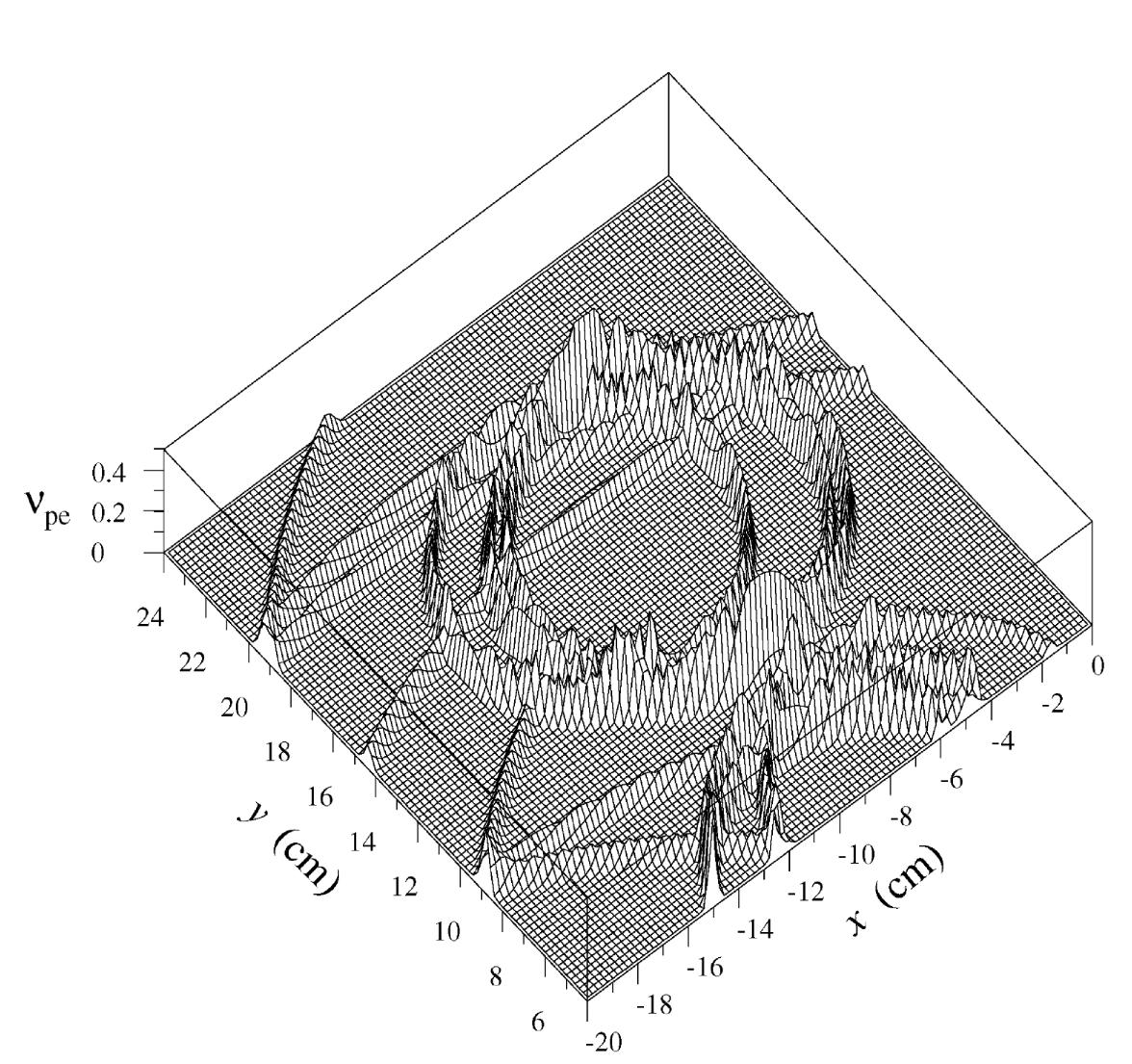}
    \caption{Left: Simulated event containing $B^0\to \pi^+\pi^-$  decay in LHCb RICH-1. Right: Expected number of photons in each pixel for the region of the left picture for a set of hypotheses that maximizes the global event PID likelihood \cite{lhcb_global}.}
    \label{fig:lhcb}
\end{figure}

In practice, the global likelihood is maximized iteratively, as an exhaustive evaluation of all possible combinations of particle hypotheses is computationally prohibitive. The procedure typically starts from an initial hypothesis assignment that reflects the expected particle composition of the experimental environment. For example, in the LHCb experiment, where the majority of charged particles produced in proton–proton collisions are pions, the initial configuration assumes the pion hypothesis for all tracks. Individual track hypotheses are then updated sequentially, and changes are accepted if they increase the overall event likelihood. This procedure is repeated until convergence is reached, i.e. no further improvement of the global likelihood is observed. In the LHCb RICH with average of about 50 tracks per event such global likelihood maximization typically takes about 5000 steps, which is manageable. An example of simulated event in LHCb RICH-1 is shown in Figure \ref{fig:lhcb}, along with the calculated expected number of photons per pixel for a given set of particle hypotheses \cite{lhcb_global}. Global likelihood methods are particularly effective when background contributions are dominated by overlapping Cherenkov rings from multiple tracks. In contrast, in scenarios where detector noise or uncorrelated photon backgrounds dominate, local likelihood approaches with an appropriate background description may provide comparable performance at much lower computational cost.

\subsection{Hough transform techniques}

The Hough transform provides an alternative pattern-recognition strategy for identifying Cherenkov rings that does not rely on external tracking information or explicit likelihood construction. In this approach, a Cherenkov ring in the photon detector plane is described by its center $(x_0,y_0)$ and radius $r$. For a detected photon hit at position $(x_i,y_i)$, the ring parameters satisfy the implicit relation
\begin{equation}
f(x_i,y_i; x_0,y_0,r) = (x_i - x_0)^2 + (y_i - y_0)^2 - r^2 = 0,
\end{equation}
which defines a two-dimensional surface in the three-dimensional Hough parameter space $H(x_0,y_0,r)$. A physical Cherenkov ring corresponds to the intersection of such surfaces arising from multiple hits, producing a localized accumulation in the discretized parameter space. The illustrated method can be generalized to treat also the elliptic hit patterns. 

In practical implementations, the Hough space is discretized and filled by accumulating contributions from all detected hits. Peaks in the accumulator identify candidate ring parameters, which can subsequently be refined using local fits or evaluated with likelihood-based methods. Variants such as weighted Hough transforms assign hit-dependent weights based on signal probabilities or detector response, improving robustness against background and detector inefficiencies. It has been successfully applied in several experiments (e.g. at ALICE \cite{alice} or CBM \cite{cbm_hough, cbm_ml} experiments), often as an initial ring-finding step followed by a refined fit using likelihood or other optimization-based methods. 

\section{Machine learning approaches}

In recent years, machine learning techniques have become an increasingly important component of particle identification and reconstruction workflows in modern experiments. The growing availability of large datasets, combined with advances in computing hardware and algorithm development, has enabled the application of ML methods to problems that were traditionally addressed using analytical or likelihood-based approaches. In the context of RICH detectors, machine learning is primarily employed to capture complex correlations between detector responses, track kinematics, and event-level properties that are difficult to model explicitly. Rather than replacing established reconstruction algorithms, ML techniques are most commonly used to complement them, either by combining information from multiple subdetectors in a global PID framework or by providing fast, high-performance alternatives for specific reconstruction tasks.

\subsection{Global particle identification with machine learning}

Machine learning methods are most widely adopted in RICH-based particle identification within the context of experiment-wide global PID frameworks, where information from several detector subsystems is combined to determine particle species. In this approach, subdetector responses such as RICH likelihoods, tracking observables, calorimetric information, and time-of-flight measurements are treated as input features to a multivariate classifier. By construction, such classifiers are able to capture non-linear correlations between observables and to incorporate additional high-level information that may be difficult to include consistently in analytical likelihood models.

Machine learning classifiers such as boosted decision trees and deep neural networks provide a flexible framework to capture these correlations and to exploit additional high-level observables. Studies in experiments such as Belle~II \cite{belle2_ml}, LHCb \cite{lhcb_ml}, and ALICE \cite{alice_ml} have demonstrated that ML-based global PID can improve efficiency and purity compared to traditional likelihood combinations, particularly in momentum regions where detector responses overlap. 

\subsubsection{Example of Belle II global PID}

A concrete example of the advantage of ML-based global PID is observed at Belle~II in the presence of decayed or strongly scattered particles traversing the ARICH detector. In such cases, the reconstructed track parameters do not accurately describe the particle trajectory inside the radiator, leading to distorted Cherenkov patterns and poorly modeled ARICH likelihood values. Within the nominal global PID, where subdetector likelihoods are combined in a factorized manner, these tracks tend to remain strongly misidentified because the ARICH likelihood enters with full weight even when its underlying assumptions are violated.

\begin{figure}[!ht]
    \centering
    \includegraphics[width=0.9\linewidth]{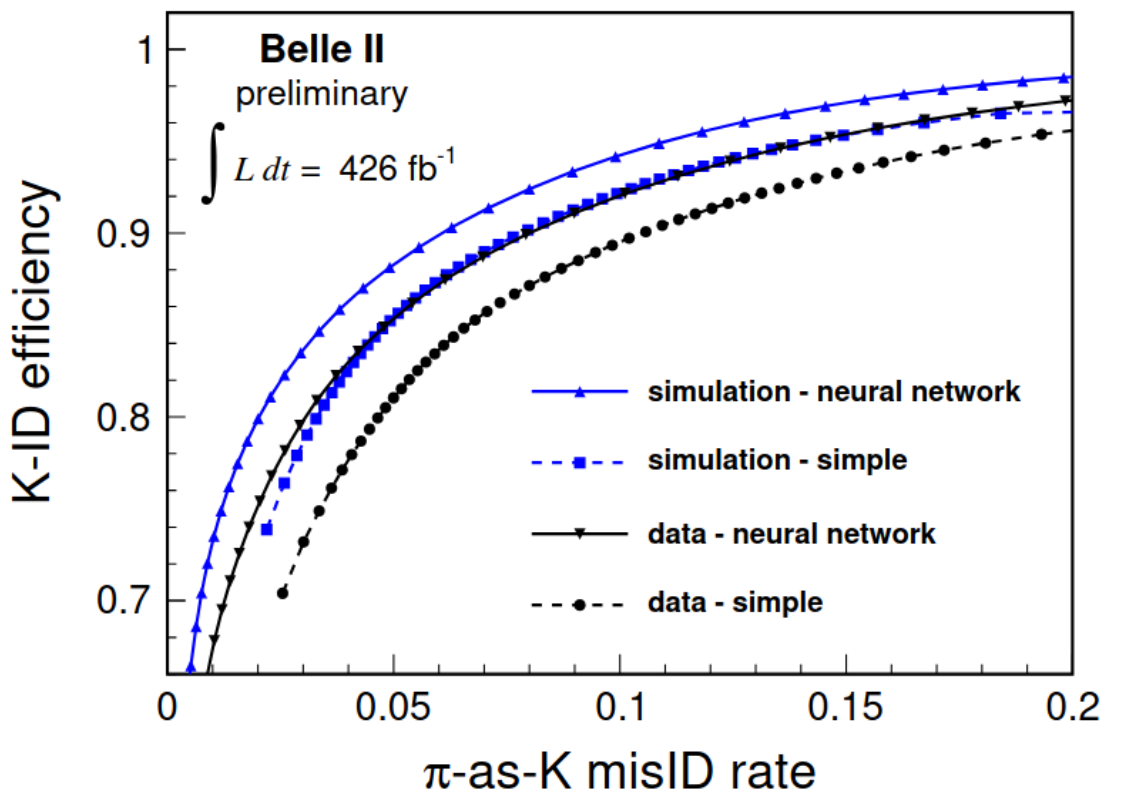}
    \caption{Global PID performance for kaon/pion separation at Belle for kaons and pions from $D^* \to (D\to K\pi)\pi$ decays. In both, simulated and measured data, notable improvement is observed when sub-detector PID likelihoods are combined by the neural network \cite{belle2_ml}.}
    \label{fig:belle_nn}
\end{figure}

When subdetector likelihoods are combined using a neural-network classifier, this effect is significantly reduced. The network learns phase-space regions in which ARICH-based likelihoods become less reliable and effectively down-weights their contribution relative to other PID observables. As can be seen in Figure \ref{fig:belle_nn} this results in notably improved global PID performance. Similar levels of PID performance improvements are reported by other experiments as referenced earlier.

\subsection{Machine learning for RICH ring reconstruction}

Machine learning techniques applied directly to RICH pattern recognition can broadly be grouped into two conceptual approaches. The first builds on established reconstruction chains by constructing higher-level observables from Cherenkov ring images and subsequently combining them using multivariate classifiers. The second, more radical approach bypasses intermediate feature engineering and feeds “raw” or minimally processed hit maps directly into neural networks.

In the first approach, geometrical or likelihood-based reconstruction methods are used to extract physics-motivated observables, such as reconstructed Cherenkov angles, ring radii, photon yields, or quality variables characterizing ring fits. These quantities are then provided as input to multivariate classifiers, such as boosted decision trees or neural networks. This strategy preserves much of the interpretability of traditional reconstruction while allowing non-linear correlations between observables to be exploited. An example of this approach is found in the CBM experiment \cite{cbm_ring_ml}, where reconstructed ring parameters and related variables are combined using machine learning techniques to improve separation power while retaining a clear connection to the underlying detector response.

\begin{figure}[!ht]
    \centering
    \includegraphics[width=1\linewidth]{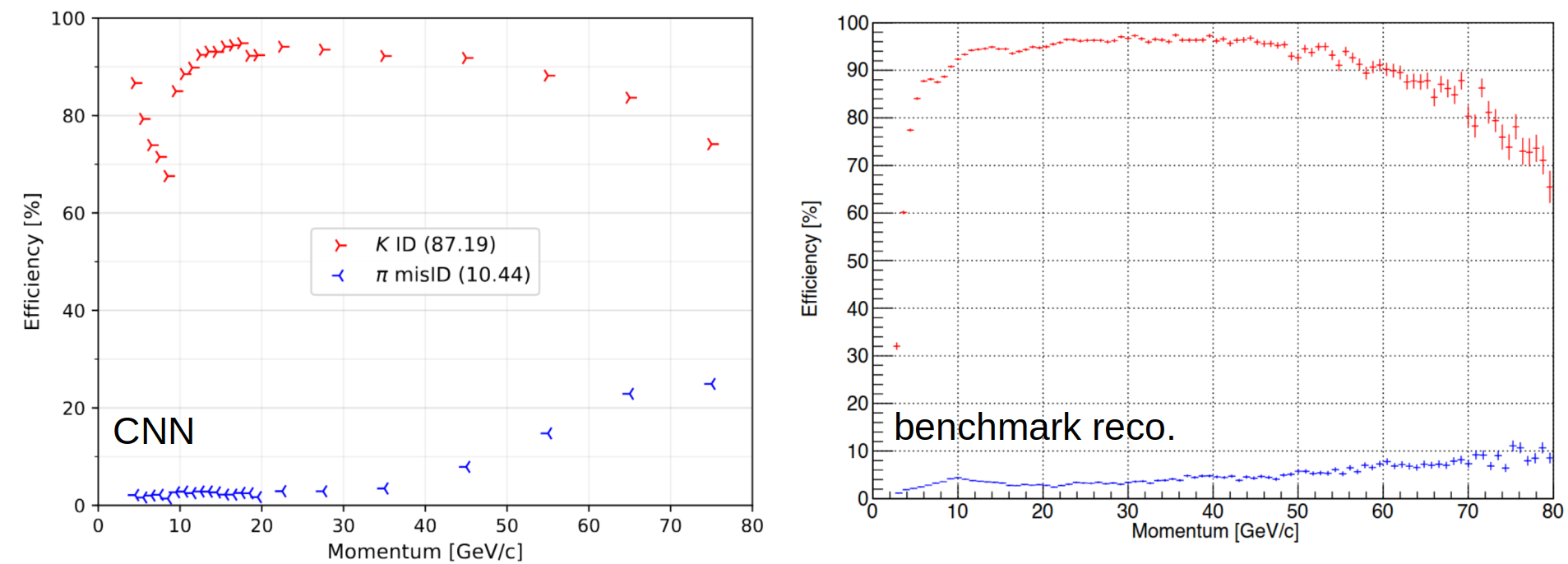}
    \caption{Comparison of kaon/pion separation vs. particle momentum at the LHCb experiment for the CNN (left) and classical likelihood (right) based reconstruction \cite{lhcb_ring_ml}.} 
    \label{fig:lhcb_cnn}
\end{figure}

\begin{figure}[!ht]
    \centering
    \includegraphics[width=1\linewidth]{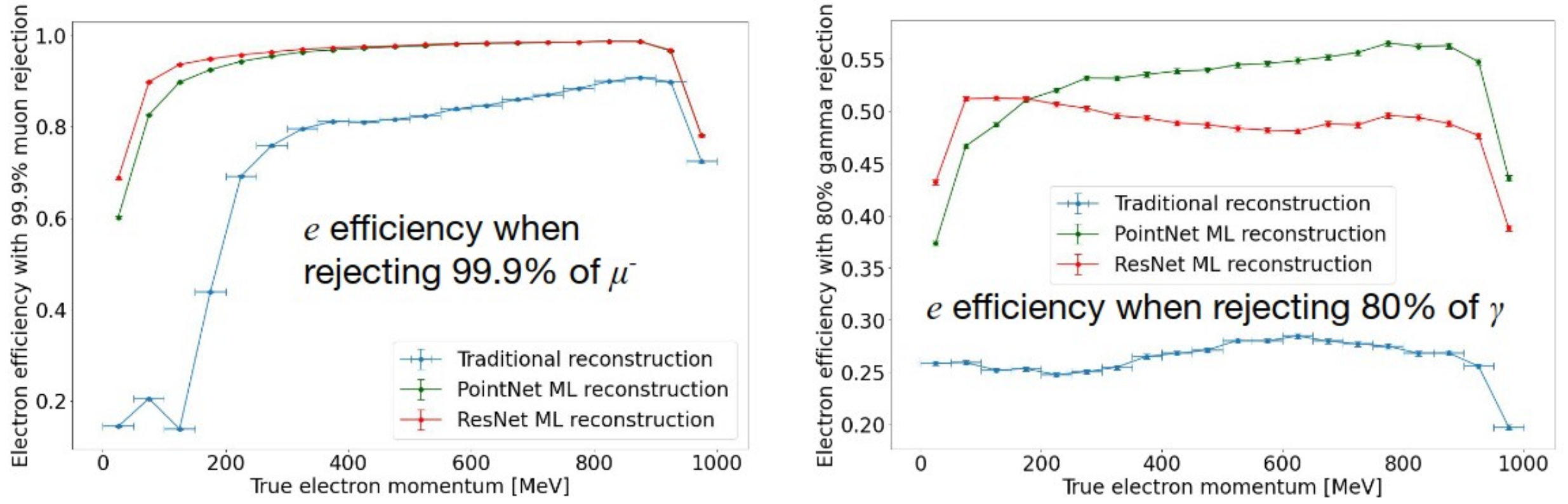}
    \caption{Left: Comparison of electron/muon (left) and electron/gamma (right) separation performance in Hyper-Kamiokande for three different reconstruction methods, two CNN based and traditional \cite{hyper_ring_ml}.}
    \label{fig:hyperk_ccn}
\end{figure}

A more direct strategy treats the Cherenkov detector output as an image-recognition problem. In this case, two-dimensional hit maps—optionally including timing information—are transformed into fixed-size representations and used as input to convolutional neural networks or related architectures. In these approaches, feature extraction is performed implicitly by the network, enabling complex and highly non-linear pattern recognition at the cost of reduced transparency compared to classical reconstruction chains. Such methods have been explored, for example, in the LHCb RICH detectors, where polar-transformed hit images are processed by deep neural networks for particle identification \cite{lhcb_ring_ml}. The obtained kaon/pion separation performance is shown in Figure \ref{fig:lhcb_cnn}, along with the performance of traditional reconstruction algorithms. While in some regions of particle momentum phase space the CNN based reconstruction achieves comparable performance, traditional likelihood based approach outperforms it most notably at high momenta, where the performance is driven by the Cherenkov angle resolution. Similar concepts are applied in large water Cherenkov detectors, where full photomultiplier hit patterns are analyzed using deep learning techniques to distinguish between different particle topologies. With increasing numbers of readout pixels in large water Cherenkov detectors (e.g. at Hyper-Kamiokande), the computational cost of traditional likelihood-based reconstruction algorithms becomes a limiting factor. ML-based approaches offer substantial gains in computational speed once trained, which makes them particularly attractive for online reconstruction and high-throughput environments. Figure \ref{fig:hyperk_ccn} shows comparison of electron/muon and electron/gamma separation performance for two CNN based algorithms with the traditional likelihood based reconstruction as obtained form the recent Hyper-Kamiokande studies \cite{hyper_ring_ml}. In both cases significant improvements in performance are observed. Once trained, these networks can process over 100,000 events per minute on one GPU, representing an improvement of ﬁve orders of magnitude over traditional reconstruction on CPU. 

Nevertheless, challenges remain in controlling systematic effects, understanding model extrapolation beyond the training domain, and maintaining robustness against changes in detector conditions. As a result, ML-based ring reconstruction is often used in parallel with, rather than as a replacement for, established reconstruction methods.

\subsection{Robustness, domain adaptation, and real-time constraints}
A key limitation of ML-based reconstruction is its dependence on the quality of the training data. Differences between simulated and real detector responses, commonly referred to as domain shift, can degrade performance if not properly addressed. Techniques such as domain adaptation, adversarial training, and feature embedding have been proposed to mitigate these effects by explicitly reducing discrepancies between simulation and data.

Another practical consideration is the estimation of systematic uncertainties associated with ML-based PID. Unlike likelihood-based methods, which often provide well-defined probabilistic interpretations, ML classifiers typically require additional studies to assess their sensitivity to variations in detector calibration and modeling assumptions. 

Largely due to these considerations, most studies of ML-based Cherenkov ring reconstruction remain at the exploratory level. Nevertheless, in some cases clear advantage in terms of PID performance and computational cost compared to the classical reconstruction methods is demonstrated, which promises great potential for future developments.  

\section{Generative models for fast simulation and reconstruction}

The detailed simulation of Cherenkov detectors is computationally demanding due to the modeling of optical photon production, propagation, and detection. In high-luminosity environments, the CPU cost of full Geant-based simulation can become a limiting factor for large-scale Monte Carlo production and detector-optimization studies. Generative machine learning models provide an alternative strategy by learning probabilistic descriptions of detector responses directly from detailed simulation or calibration data.

\begin{figure}[!ht]
    \centering
    \includegraphics[width=1\linewidth]{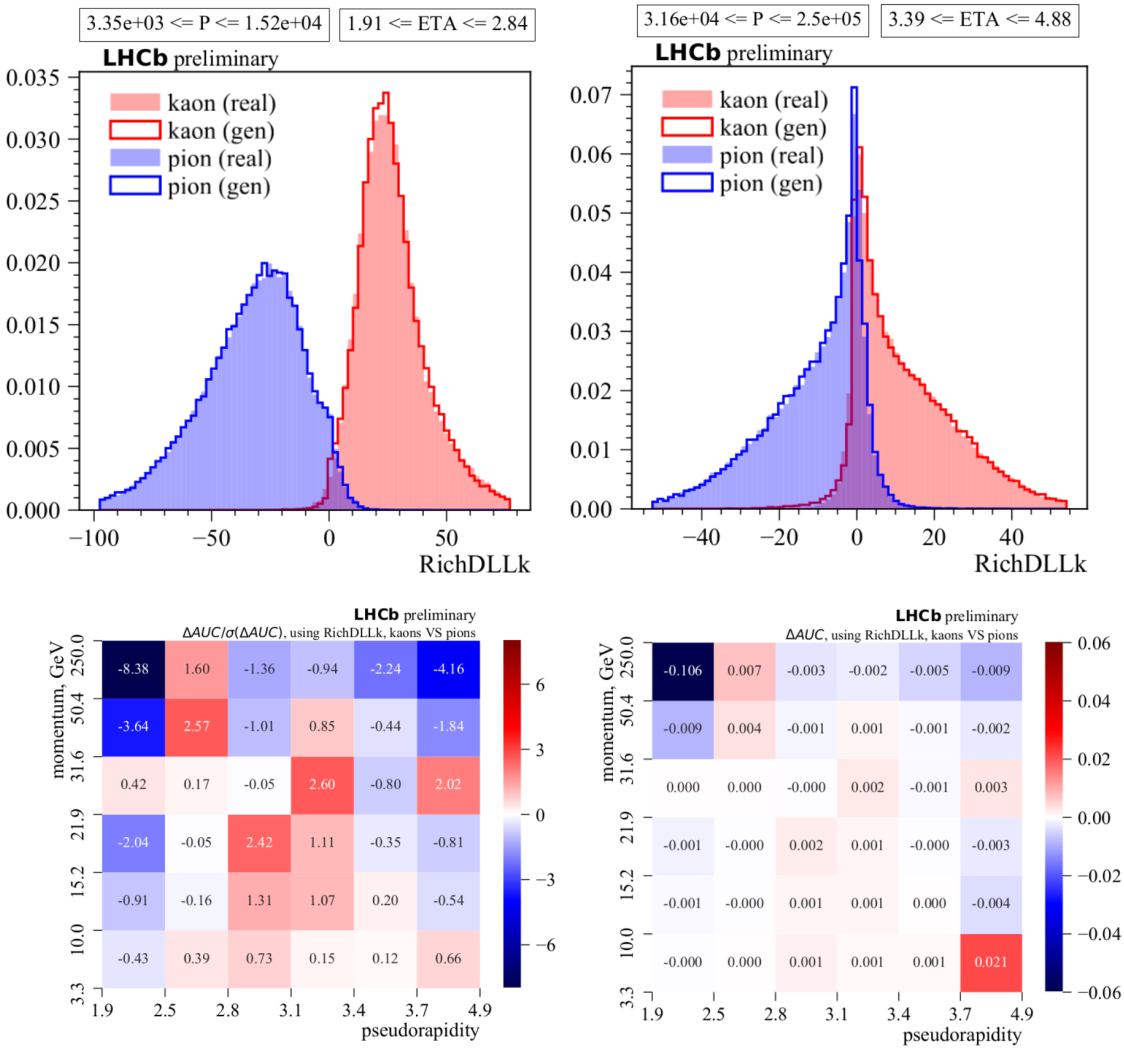}
    \caption{Upper: Comparison of the log-likelihood difference between the kaon and pion hypotheses for pion and kaon samples in LHCb data and as generated by the GAN-based model, for two kinematic regions (indicated above the plots). Bottom: Significance (left) and absolute values (right) of the observed differences in the corresponding areas under the ROC curve (AUC), in bins of particle momentum and pseudorapidity \cite{lhcb_gan}.}
    \label{fig:lhcb_gan}
\end{figure}

A first class of approaches focuses on the generation of high-level reconstructed observables rather than low-level detector hits. A representative example is provided by the LHCb experiment, where generative adversarial networks (GANs) are trained to reproduce PID likelihood differences (e.g.\ RichDLL quantities) as functions of track kinematics and event occupancy \cite{lhcb_gan}. In this approach, the model learns the multidimensional distribution of reconstructed PID observables directly from calibration samples, without explicitly simulating individual photons or optical transport. The trained generator produces statistically consistent likelihood distributions at a fraction of the computational cost of full detector simulation, making this strategy particularly attractive for large-scale Monte Carlo production and systematic studies requiring substantial simulated statistics. Figure \ref{fig:lhcb_gan} shows comparison of PID likelihood distributions for pions and kaons as observed in real data and as generated by the GAN model for two different regions of track kinematic phase-space. Apart from the regions of phase-space with sparsely populated training data, very good agreement in distributions and resulting PID performance is observed.

A conceptually different direction aims to model detector responses at the hit-pattern level. Architectures such as DeepRICH employ generative models based on variational autoencoders (VAEs) to learn compact representations of Cherenkov photon distributions \cite{deep_rich, fanelli_2025, fanelli_2025_1}. Photon hits—described by spatial coordinates and, where available, timing information—are processed together with track kinematic parameters by an encoder network that maps them into a structured latent space. A decoder network then generates hit patterns from this latent representation, and the model is trained by minimizing a reconstruction loss supplemented by regularization terms that enforce a smooth and stable latent distribution.

When particle type is introduced, it is typically used as a conditioning variable rather than as a prediction target. The model therefore learns the conditional probability density
\[
p(\text{hits} \mid \text{kinematics}, h),
\]
providing a flexible, data-driven parameterization of detector response for different particle hypotheses. This generative objective differs fundamentally from supervised classification, where the task is to infer the particle type from observed hits. While some implementations augment the architecture with a dedicated classification branch, the generative component itself is trained to model detector-level response distributions.

Such detector-level generative models can be used in two distinct but related contexts. First, they can act as fast simulators: once trained, the decoder can generate statistically consistent Cherenkov hit patterns conditioned on track parameters, replacing computationally expensive photon transport in large-scale Monte Carlo production. Second, the learned generative model can be incorporated directly into the reconstruction chain, providing an effective and differentiable parameterization of detector response functions or probability density distributions entering likelihood-based PID. In this setting, the emphasis lies not only on computational efficiency but also on accurately capturing complex spatial and temporal correlations—such as reflections, chromatic dispersion, or detector granularity—that are difficult to describe analytically.

\begin{figure}[!ht]
    \centering
    \includegraphics[width=0.8\linewidth]{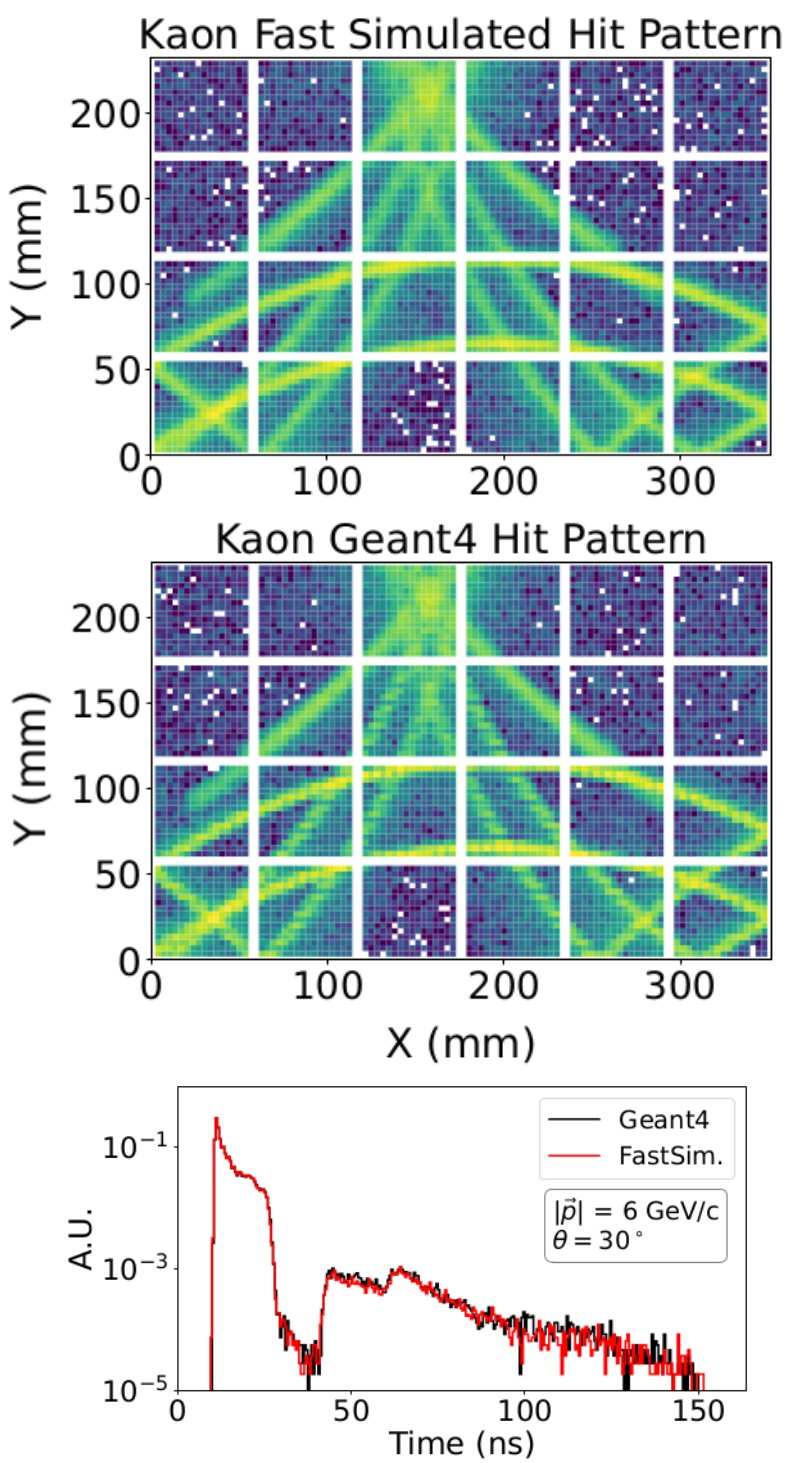}
    \caption{Comparison of the photon hit patterns for 6 GeV kaons at the future Electron-Ion Collider High-Performance DIRC as generated by the Geant4 simulation and by the GAN based fast simulation. Upper two plots compare hits spatial distributions while the bottom plot compares distributions in photon time arrival \cite{fanelli_2025_1}.}
    \label{fig:deep_rich_sim}
\end{figure}

\begin{figure}[!ht]
    \centering
    \includegraphics[width=0.9\linewidth]{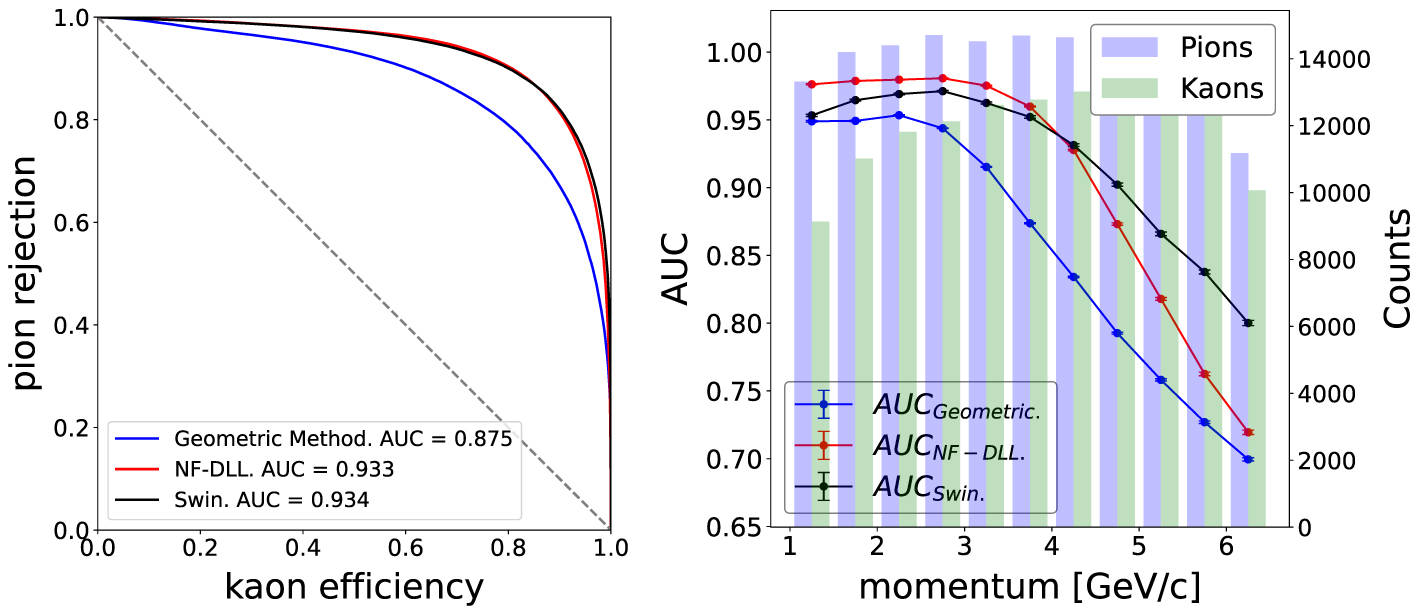}
    \caption{Left: Comparison of pion/kaon ROC curves for the GlueX DIRC detector as obtained by the use of standard geometric reconstruction and two GAN-based likelihood evaluation methods (Swin architecture - in black and Normalizing Flow method - in red), integrated over the entire phase space. Right: AUC as a function of track momentum for all three methods \cite{fanelli_2025}.}
    \label{fig:deep_rich_reco}
\end{figure}

Recent developments extend this paradigm using conditional normalizing flows and diffusion-based models. Normalizing flows provide invertible mappings with tractable likelihood evaluation, enabling direct integration into likelihood-based reconstruction frameworks. Diffusion models offer highly flexible generative mechanisms capable of describing complex, high-dimensional Cherenkov photon distributions across broad regions of phase space. Applications to DIRC and RICH detectors at future facilities demonstrate that such approaches can reproduce Cherenkov angle distributions, photon yields, and PID observables with high fidelity, while maintaining substantial computational advantages. Figure \ref{fig:deep_rich_sim} compares the photon hit patterns for 6 GeV kaons at the future Electron-Ion Collider High-Performance DIRC as generated by the Geant4 simulation and by the GAN based fast simulation. Very good agreement between the complex distributions is observed \cite{fanelli_2025_1}. Figure \ref{fig:deep_rich_reco}, on the other hand, shows the comparison of pion/kaon separation power (ROC curve) for the GlueX DIRC detector as obtained by the use of standard geometric reconstruction and GAN-based likelihood evaluation methods \cite{fanelli_2025}. The latter clearly provide notable improvement in the PID performance.

Although generative models offer impressive performance and significant speed gains, they do not eliminate the need for detailed detector simulation. High-quality simulation or calibration data remain essential for training and validation, and careful treatment of systematic uncertainties is required, particularly in sparsely populated regions of phase space. Generative methods are therefore best viewed as powerful complements to traditional simulation and reconstruction techniques, enabling efficient, probabilistically consistent modeling of Cherenkov detector responses.

\section{Conclusions}

Traditional reconstruction methods, in particular likelihood-based approaches and Hough-transform techniques, continue to form the backbone of RICH data analysis in many experiments. At the same time, the rapid development of machine learning methods has opened new possibilities for global particle identification, ring reconstruction, and fast simulation.

While machine learning offers clear advantages in terms of speed and, in some cases, performance, these tools should be used with care and a clear understanding of their limitations. In many situations, traditional methods remain optimal for well-defined fitting problems, while machine learning excels in handling complex correlations and high-dimensional input spaces. The continued interplay between established reconstruction techniques and modern machine learning approaches is expected to drive further improvements in RICH detector performance, particularly in view of upcoming high-luminosity and high-granularity detectors.

\section*{Acknowledgements}
The author is supported by the Slovenian Research and Innovation Agency research grants No. J1-50010 and P1-0135.

\bibliographystyle{elsarticle-num}

\begin{thebibliography}{99}
\bibitem{likeli} P. Baillon, Nucl. Instrum. Meth. A, 238:341, 1985.
\bibitem{belle2_arich1} L. \v Santelj et al., Nucl. Instrum. Meth. A, 876:104-107, 2017.
\bibitem{belle2_arich2} L. \v Santelj et al., Nucl. Instrum. Meth. A, 1055:168502, 2023.
\bibitem{lhcb_global} R. Forty et al., Nucl. Instrum. Meth. A, 433:257-261, 1999.
\bibitem{alice} B. Alessandro, et al., (ALICE Collaboration) J. Phys. G: Nucl. Part. Phys., 32, p. 1295, 2006. 
\bibitem{cbm_hough} J. Adamczewski et al., Nucl. Instrum. Meth. A, 766: 250-254, 2014.
\bibitem{cbm_ml} M. Beyer, EPJ Web of Conferences 337, 01248, 2025.
\bibitem{belle2_ml} Belle II Collaboration, Eur. Phys. J. C 85, 1237, 2025.
\bibitem{lhcb_ml} D. Derkach et al., J. Phys.: Conf. Ser. 1085, 042038, 2018.
\bibitem{alice_ml} M. Karwowska et al., EPJ Web of Conferences 295, 09029 (2024)
\bibitem{cbm_ring_ml} H. Schiller, Application of Machine Learning to Particle Identification for Dielectron Analysis in CBM [doctoral dissertation, M\"unster university], 2022.
\bibitem{lhcb_ring_ml} M. P. Blago, J. Phys.: Conf. Ser. 2438 012076, 2023.
\bibitem{hyper_ring_ml} N. Prouse et al., Phys. Sci. Forum, 8(1), 63., 2023. 
\bibitem{lhcb_gan} A Maevskiy et al., J. Phys.: Conf. Ser. 1525 012097, 2020.
\bibitem{deep_rich} C. Fanelli et al., Mach. Learn.: Sci. Technol. 1 015010, 2020.
\bibitem{fanelli_2025} C. Fanelli et al., Mach. Learn.: Sci. Technol. 6 015028, 2025.
\bibitem{fanelli_2025_1} J. Giroux, arXiv:2504.19042 [physics.ins-det], 2025.

\end{thebibliography}

\end{document}